\documentclass[runningheads]{svmult}

\usepackage{makeidx}   
\usepackage{graphicx}  
\usepackage{subeqnar}  
\usepackage{multicol}  
\usepackage{physprbb}  
\makeindex             

\def\la{\raisebox{-0.5ex}{$\,\stackrel{<}{\scriptstyle\sim}\,$}}
\def\ga{\raisebox{-0.5ex}{$\,\stackrel{>}{\scriptstyle\sim}\,$}}

\def\etal{et al.}

\def\teff{\ifmmode T_{\rm eff} \else $T_{\mathrm{eff}}$\fi}

\def\ltsima{$\buildrel<\over\sim$}
\def\lsim{\lower.5ex\hbox{\ltsima}}

\newcommand{\ha}{\ifmmode {\rm H}\alpha \else H$\alpha$\fi}
\newcommand{\hb}{\ifmmode {\rm H}\beta \else H$\beta$\fi}
\newcommand{\lya}{Lyman-$\alpha$}
\newcommand{\hei}{He~{\sc i}}

\newcommand{\heii}{He~{\sc ii}}
\newcommand{\Heiiuv}{He~{\sc ii} $\lambda$1640}
\newcommand{\Heiiopt}{He~{\sc ii} $\lambda$4686}
\newcommand{\qh}{\ifmmode q({\rm H}) \else $q({\rm H})$\fi}
\newcommand{\qhe}{\ifmmode q({\rm He^0}) \else $q({\rm He^0})$\fi}
\newcommand{\qhep}{\ifmmode q({\rm He^+}) \else $q({\rm He^+})$\fi}
\newcommand{\Qh}{\ifmmode Q({\rm H}) \else $Q({\rm H})$\fi}
\newcommand{\Qhe}{\ifmmode Q({\rm He^0}) \else $Q({\rm He^0})$\fi}
\newcommand{\Qhep}{\ifmmode Q({\rm He^+}) \else $Q({\rm He^+})$\fi}
\newcommand{\Qhtwo}{\ifmmode Q({\rm LW}) \else $Q({\rm LW})$\fi}
\newcommand{\qrathe}{\ifmmode q({\rm He^0})/q({\rm H}) \else $q({\rm He^0})/q({\rm H})$\fi}
\newcommand{\qrathep}{\ifmmode q({\rm He^+})/q({\rm H}) \else $q({\rm He^+})/q({\rm H})$\fi}
\newcommand{\Qrathe}{\ifmmode Q({\rm He^0})/Q({\rm H}) \else $Q({\rm He^0})/Q({\rm H})$\fi}
\newcommand{\Qrathep}{\ifmmode Q({\rm He^+})/Q({\rm H}) \else $Q({\rm He^+})/Q({\rm H})$\fi}
\newcommand{\Qhave}{\ifmmode \bar{Q}({\rm H}) \else $\bar{Q}({\rm H})$\fi}
\newcommand{\Qheave}{\ifmmode \bar{Q}({\rm He^0}) \else $\bar{Q}({\rm He^0})$\fi}
\newcommand{\Qhepave}{\ifmmode \bar{Q}({\rm He^+}) \else $\bar{Q}({\rm He^+})$\fi}
\newcommand{\Qhtwoave}{\ifmmode \bar{Q}({\rm H}_2) \else $\bar{Q}({\rm H}_2)$\fi}
\newcommand{\Qratheave}{\ifmmode \bar{Q}({\rm He^0})/\bar{Q}({\rm H}) \else $\bar{Q}({\rm He^0})/\bar{Q}({\rm H})$\fi}
\newcommand{\Qrathepave}{\ifmmode \bar{Q}({\rm He^+})/\bar{Q}({\rm H}) \else $\bar{Q}({\rm He^+})/\bar{Q}({\rm H})$\fi}

\def\msun{\ifmmode M_{\odot} \else M$_{\odot}$\fi}
\def\zsun{\ifmmode Z_{\odot} \else Z$_{\odot}$\fi}
\def\lsun{\ifmmode L_{\odot} \else L$_{\odot}$\fi}

\def\mup{\ifmmode M_{\rm up} \else M$_{\rm up}$\fi}
\def\mlow{\ifmmode M_{\rm low} \else M$_{\rm low}$\fi}

\def\aap{A\&A}

\def\apj{ApJ}

\def\mnras{MNRAS}

\usepackage{psfig}   
\begin{document}
\title*{Observing the first galaxies  --- a case for an intermediate
resolution multi-object IR spectrograph}
%
%
%
%
\titlerunning{Observing the first galaxies with the VLT}
%
\author{Daniel Schaerer\inst{1}
\and Roser Pell\'o\inst{1}}
%
%
%
\institute{Observatoire Midi-Pyr\'en\'ees, Laboratoire d'Astrophysique, UMR
 5572, 14, Av.  E. Belin, F-31400 Toulouse, France }

\maketitle              

\begin{abstract}
We present a special science case for an intermediate resolution
multi-object IR spectrograph for the VLT.
We have constructed new models for massive Population III stars
and metal-free stellar populations (see Schaerer 2001). 
The properties of individual stars and integrated populations, 
including their ionising fluxes and SEDs are discussed. 
We also study their dependence on the poorly known IMF 
and the star formation history. 
The synthetic spectra are used to simulate spectroscopic observations
of the expected emission lines of Pop III galaxies.
We show that such an instrument should allow the detection and 
efficient observations of the first galaxies with the VLT.

\end{abstract}

\section{Introduction}
\label{s_intro}

Important advances have been made in recent years on the modeling
of the first stars and galaxies forming out of pristine matter 
--- so called Pop III objects ---
in the early Universe (cf.\ review of Loeb \& Barkana 2001, proceedings 
of Weiss \etal\ 2000 and Umemura \& Susa 2001).
With no doubt these efforts are motivated by the approaching 
possibility of direct observations of such objects at very high redshift 
with NGST and large ground-based telescopes.
 
{\em What are the expected observational signatures of Pop III galaxies ?}
Tumlinson \& Shull (2000, hereafter TS00) have recently pointed out 
the exceptionally strong He$^+$ ionising flux of massive ($M \ga$ 40 \msun) 
Pop III stars, which must be a natural consequence of their compactness, 
i.e.\ high effective temperatures, and non-LTE effects in their atmospheres
increasing the flux in the ionising continua.
As a consequence strong \heii\ recombination lines such as \Heiiuv\ or 
\Heiiopt\ are expected; together with AGN a rather unique feature of 
metal-free stellar populations, as discussed by TS00 and Tumlinson 
\etal\ (2001, hereafter TGS01).
Instead of assuming a ``standard'' Population I like Salpeter IMF up to 
100 \msun\ as TS00, Bromm \etal\ (2001, hereafter BKL01) have considered stars with masses 
larger than 300 \msun, which may form according to some recent hydrodynamical
models.
An even stronger ionising flux and stronger H and \heii\ emission lines
were found.

Some strong simplifying assumptions are, however, made in the calculations of 
TS00, TGS01, and BKL01.
\begin{itemize}
\item[{\em 1)}] All stars are assumed to be on the zero age main sequence, 
   i.e.\ stellar evolution proceeding generally to cooler temperatures is neglected
   (but cf.\ TGS01 for a simple estimate of this effect).
\item[{\em 2)}] BKL01 take the hardness of the stellar spectrum of 
   the hottest (1000 \msun) star as representative for all stars with masses down 
   to 300 \msun. This leads in particular to an overestimate of the \heii\
   recombination line luminosities.
\item[{\em 3)}] None of the studies include nebular continuous emission, which 
   cannot be neglected for metal-poor objects with such strong ionising fluxes.
   This process increases significantly the total continuum flux at wavelengths 
   redward 
   of Lyman-$\alpha$ and leads in turn to reduced emission line equivalent widths.
\item[{\em 4)}] A single, fixed, IMF is considered only by both groups.
   In view of the uncertainties on this quantity it appears useful to explore
   a wider range of IMFs.
\end{itemize}

We here present results from new calculations which 
relax all the above assumptions and allow
the exploration of a wide parameter space in terms of stellar tracks 
(including also alternate tracks with strong mass loss), 
IMF, and star formation history (instantaneous bursts or constant 
SFR). A more detailed account is given in Schaerer (2001).
The models are then used to simulate observations with a medium resolution
cryogenic IR spectrograph for the VLT.

\vspace*{-0.2cm}
\section{Input physics and results for individual stars}
To account for non-LTE effects, which are crucial for a proper description
of the spectra of hot stars considered here we use the {\em TLUSTY} 
code of Hubeny \& Lanz (1995). 
Several additional models were
calculated with the {\em CMFGEN} code of Hillier \& Miller (1998)
to explore the possible importance of mass loss (as well known for Pop 
I stars, Schaerer \& de Koter 1997) on the ionising spectra of Pop III stars.
For the hottest stars (\teff $\ga$ 80 kK) dominating especially the 
ionising spectrum in the \heii\ continuum, no significant differences
with plane parallel models are found
as their continua are formed deep in the quasi static part of the atmosphere.

We use stellar tracks from 1 to 500 \msun\ calculated with the Geneva 
stellar evolution code (Desjacques 2000) or tracks from Marigo \etal\ (2001).
Although probably unrealistic, we also explore the impact of high mass loss
tracks from Klapp (1983) and El Eid \etal\ (1983).
The atmosphere models and tracks are included in the synthesis code
of Schaerer \& Vacca (1998), which also calculates nebular line and
continuous emission (adapted here to include additional lines and 
for an appropriate electron temperature $T_e=30000$ K).

Based on these models we have calculated the ionising properties
of ZAMS stars and their time averaged properties taking into account
their evolution. The quantities are tabulated and given as 
fit formulae of use for various calculations (see Schaerer 2001).

\vspace*{-0.4cm}
\section{Properties of Pop III ``galaxies''}

\begin{figure*}[tb]
\centerline{\psfig{file=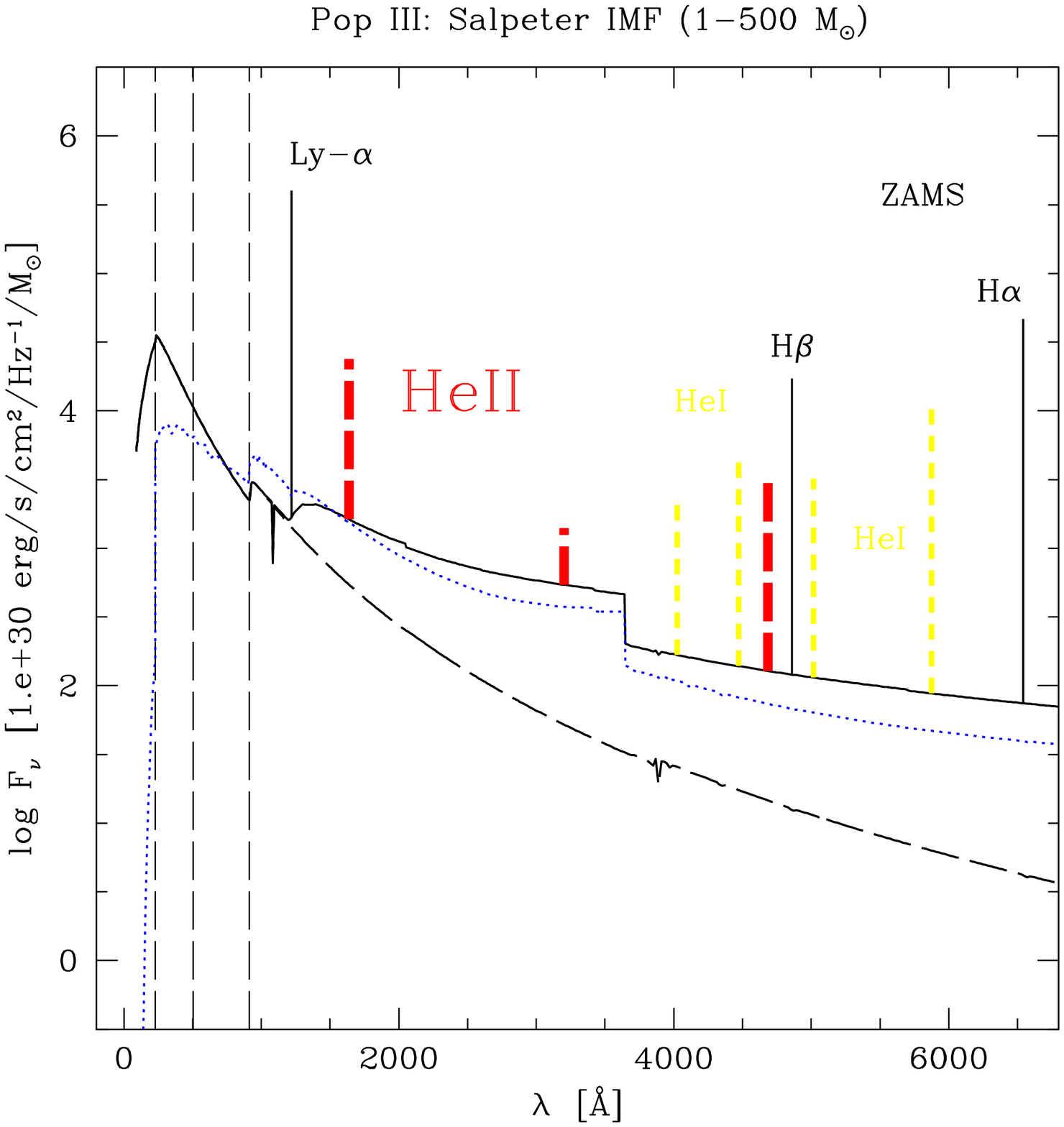,width=7cm}
  \psfig{file=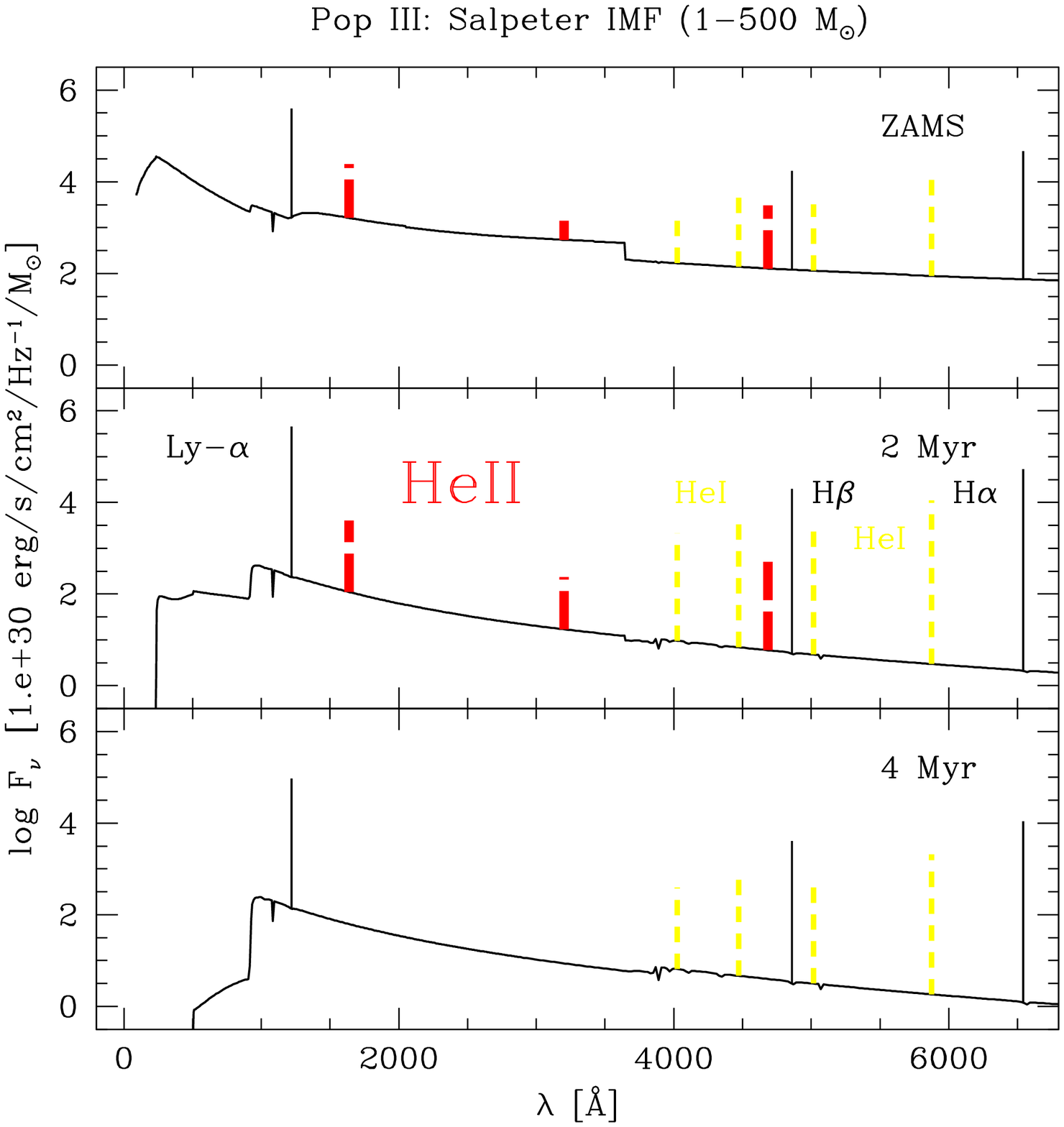,width=7cm}}
\caption{Spectral energy distribution (SED) including H and He recombination lines
for Salpeter IMF from 1 to 500 \msun.
%
{\bf Left panel:} ZAMS population.
The pure stellar continuum (neglecting nebular emission) is shown by the
dashed line. For comparison the SED of the $Z=1/50 \zsun$ population
(Salpeter IMF from 1 -- 150 \msun) is shown by the dotted line.
The vertical dashed lines indicate the ionisation potentials of H, He$^0$, and 
He$^+$. 
Note the presence of the unique \heii\ features (shown as thick dashed lines)
and the importance of nebular continuous emission.
{\bf Right panel:} temporal evolution of the spectrum for 0, 2 and 4 Myr
showing the rapid change of the emission line spectrum, characterised by the 
disappearance of the \heii\ lines
}
\label{fig_seds}
\end{figure*}

Integrated spectra have been calculated for a variety of IMFs.
SEDs of integrated zero metallicity stellar
populations are shown in Fig.\ \ref{fig_seds} for the case of a Salpeter
IMF from 1 to 500 \msun\ and instantaneous bursts of ages 0 (ZAMS), 2,
and 4 Myr. Overplotted on the continuum (including stellar $+$ nebular 
emission: solid lines) are the strongest emission lines for illustration 
purpose.
%
The striking differences when compared with non-zero metallicity, 
most importantly in the ionising flux above 
the \heii\ edge ($>$ 54 eV), have already been discussed by TS00.
The comparison of the total spectrum (solid line) with the pure stellar emission
(dashed) illustrates the importance of nebular continuous emission neglected
in earlier studies (TS00, BKL01), which dominates the ZAMS 
spectrum at $\lambda \ga$ 1400 \AA.
The nebular contribution, whose emission is proportional to \Qh,
depends rather strongly on the age, IMF, and 
star formation history. For the parameter space explored (see Schaerer 2001),
we find that nebular continuous emission is 
not negligible for bursts with ages $\la$ 2 Myr and for constant star formation
models.

The spectra in Fig.\ \ref{fig_seds} show in addition to the H and \hei\ 
recombination lines 
the presence of the 
strong \heii\ $\lambda\lambda$ 1640, 3203, and 4686 recombination lines, which 
--- due to the exceptional hardness of the ionising spectrum ---
represent a unique feature of Pop III starbursts compared to metal enriched 
populations (cf.\ TGS01, Oh \etal\ 2001, BKL01).
Another effect highlighted by this Figure is the rapid temporal evolution of 
the recombination line spectrum. Indeed, already $\ga$ 3 Myr after the burst, 
the high excitation lines are absent, for the reasons discussed before.

The detailed behaviour of the relative line intensities and equivalent widths
and their dependence on the IMF and the star formation history are 
discussed in Schaerer (2001).
For the case of constant star formation, the \heii\ or H recombination
lines can be used as a measure of the star formation rate (SFR).
For obvious reasons, strong variations of the various SFR indicators
on the poorly known IMF are obtained.

\section{Simulations for a 2nd generation multi-object IR spectrograph
on the VLT}

We have used the above synthetic spectra to calculate the photometric
properties and to simulate spectroscopic observations.
The following assumptions are made:
A popular cosmology ($\Omega_m = 0.3$, $\Omega_\Lambda = 0.7$, $H_0 =
75$ km/s/Mpc$^{-1}$) is adopted.
\lya\ forest blanketing at redshift $z <6$ is included
following the prescription of Madau (1995).
Sources are unresolved on a 0.3'' scale, and the seeing adopted is 0.8''.
We take typical SOFI and ISAAC near-IR filter responses to calculate 
magnitudes in the Vega system. Telescope parameters correspond to the VLT. 
The characteristics of the near-IR spectrograph are set similar to 
EMIR (Balcells et al. 2000), with 0.3''/pixel, 1'' slit-width, and a
mean total efficiency of 40\%. 
All simulations shown here are calculated for a ZAMS population of
10$^7$ \msun\ with a  Salpeter IMF from 1 to 500 \msun\
(somewhat more ``favourable'' than a constant star formation case),
a total exposure time of 10$^5$ sec, and resolutions
$R \sim$ 1000 to 5000.

\begin{figure*}[tb]
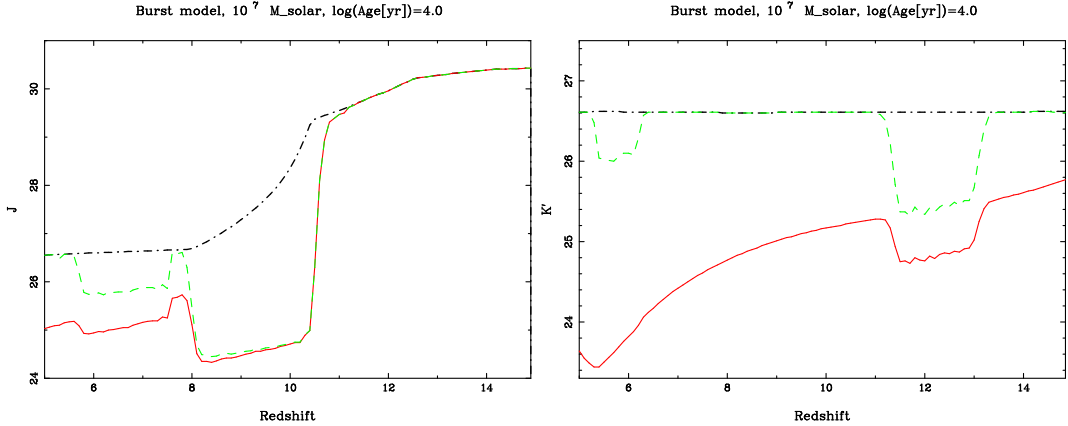

\centerline{\psfig{file=dschaerer_fig2a.eps,angle=270,width=7cm}
            \psfig{file=dschaerer_fig2b.eps,angle=270,width=7cm}}
\caption{J (left) and K$^\prime$ (right) band magnitude as a function of redshift over the
interval $z \sim$ 6 to 14.5.
Pure stellar continuum: upper dashed-dotted curve; intermediate/green
dashed curve: including the \lya, \Heiiuv, and He~{\sc ii} $\lambda$3203
emission lines; total predicted spectrum including lines
and nebular continuous emission (lower curve/red). Note the increase
by $\sim$ 1 -- 2.5 mag depending on $z$!}
\label{fig_k}
\end{figure*}

\begin{figure*}[tb]
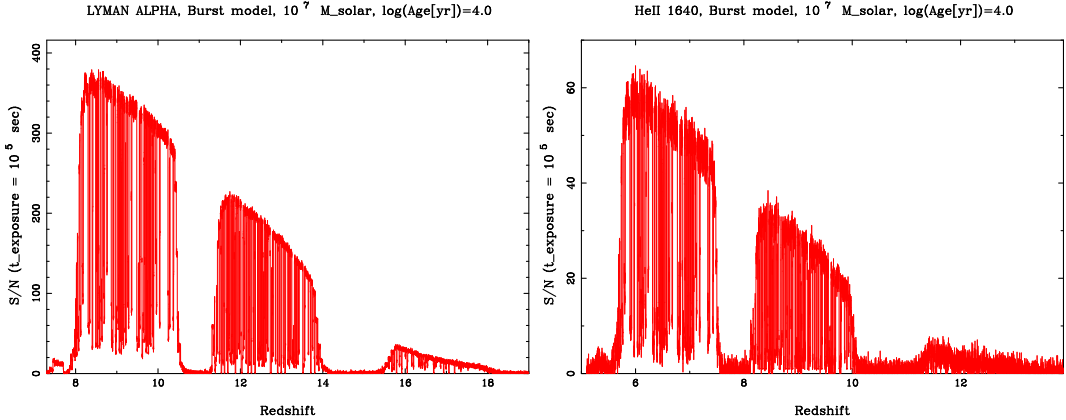

\centerline{\psfig{file=dschaerer_fig3a.eps,angle=270,width=7cm}
            \psfig{file=dschaerer_fig3b.eps,angle=270,width=7cm}}
\caption{Predicted S/N in \lya\ (left) and \Heiiuv\ (right) as a
            function $z$ for a spectral resolution $R=1000$
            and $t_{\rm exp} = 10^5$ s, for a $10^7$ \msun\ burst
            at age $\sim$ 0 with a Salpeter IMF from 1 to 500 \msun}
\label{fig_snr}
\end{figure*}

As shown in Fig.\ \ref{fig_k} the predicted magnitude in the K$^\prime$ band
is $\sim$ 24 -- 25 for a 10$^7$ \msun\ burst. 
This Fig.\ shows also the strong contribution of the nebular continuum,
and the signatures of the strong emission lines on this broad band
filter, which both need to be taken into account. Similar effects are 
observed in the other filter bands, the most relevant emission lines being 
\lya\ and \Heiiuv.
Simulations for various cases of the IMF show that standard broad-band colors 
do not allow a distinction between upper mass cut-offs of 100 to 1000
\msun\ and are also insensitive to changes of \mlow.

The expected S/N in \lya\ and \Heiiuv\ as a function of redshift
of the source (observed in the JHK bands) for a spectral resolution of 1000 is 
plotted in Fig.\ \ref{fig_snr}. 
The calculations illustrate the following:
\begin{itemize}
\item \lya\ can easily be detected with a good S/N over a redshift
ranges ($z \sim$ 8--10.5, 11.5--14, and 15.5--18).
A joint detection with \Heiiuv\ is possible for $z \sim$ 5.5--7.5
(\lya\ in optical domain) and $z \sim$ 8--10; at larger redshift the S/N of
\Heiiuv\ becomes very low.

A detection of both \Heiiuv\ and \lya\ is obviously important to secure
the redshift, and to obtain a measure of the hardness of the ionising flux.
This allows potentially the distinction between different IMFs.

\item Higher spectral resolution ($R \sim$ 5000) considerably increases
the chances of detection between the sky lines. For unresolved lines
the resulting decrease of S/N is modest.
The medium spectral resolution is also required to attempt to measure
the emission line profiles, in order to distinguish Pop III source from
AGN (cf.\ TGS01).
Once this is obtained, the data can be rebinned to increase the S/N.
\end{itemize}

{\em How many such Pop III galaxies are expected ?}
The number of metal-free objects depends of course on 
the galaxy formation scenario, the star formation efficiency, and
the hydrodynamical enrichment process of the ISM, among other factors.
The comprehensive study of Ciardi \etal\ (2000) shows that
at $z \ga$ 8 naked stellar clusters, i.e.\ objects which have complete
blown out their ISM dominate the population of luminous objects.
One may expect that all metals are ejected in this way into the IGM,
thereby avoiding largely local pollution, i.e.\ leaving these objects
as Pop III ``galaxies''.
Based on such an assumption and using the ionising fluxes from TS00, 
Oh \etal\ (2001) have recently calculated
the predicted number of Pop III starbursts detectable in \heii\ lines
with NGST for a one day integration.
Their estimate yields between $\sim$ 40 and 3000 sources in a 8'x8' 
field of view envisaged for the proposed IR multi object spectrograph.
Although pilot studies have recently started to explore the formation of dust
in Pop III objects (Todini \& Ferrara 2001), the effect is neglected here.

In conclusion, a medium resolution ($R$ up to $\sim$ 5000) IR multi-object 
spectrograph on the VLT covering the J, H, and K band would not only be 
of great interest to numerous studies of intermediate and high-redshift
galaxies (cf.\ contributions of Cuby, Genzel, Lef\`evre, this meeting).
It should also allow observations of the first so called Pop III objects
--- certainly one the great remaining challenges of observational
astronomy. 
If realised on a reasonable time scale, such ground-based observations
with the VLT at ESO should be able to achieve an important breakthrough
before the launch of NGST.


%

\end{document}